\documentclass[twocolumn,pre,showpacs,floatfix]{revtex4}
\usepackage{epsfig}
\usepackage{amsfonts}
\usepackage{amssymb}
\usepackage{amsmath}
\usepackage{pgf,pgfarrows,pgfnodes,pgfautomata,pgfheaps,pgfshade}
\usepackage{graphics,graphicx}

\newcommand{\Z}{{\mathbb{Z}}}

\begin{document}
\title{Fluxon mobility in an asymmetric SQUID array}
\author{Yaroslav Zolotaryuk}
\email{yzolo@bitp.kiev.ua}
\affiliation{Bogolyubov Institute for Theoretical Physics,
National Academy of Sciences of Ukraine,
vul. Metrologichna 14B, 
 03680 Kyiv, Ukraine}
\author{Ivan O. Starodub}
\email{starodub@bitp.kiev.ua}
\affiliation{Bogolyubov Institute for Theoretical Physics,
National Academy of Sciences of Ukraine,
vul. Metrologichna 14B, 
 03680 Kyiv, Ukraine} 

\date{\today}

\begin{abstract}
Fluxon dynamics in the dc-biased array of asymmetric three-junction 
superconducting quantum interference devices (SQUIDs) is investigated.
The array of SQUIDs is described by the discrete double sine-Gordon
equation. It appears that this equation possesses a finite set of
velocities at which the fluxon propagates with the constant shape and
 without radiation. The signatures of these velocities appear on 
the respective current-voltage characteristics of the array as 
 inaccessible voltage intervals (gaps). The critical depinning current
has a clear minimum as a function of the asymmetry parameter (the
ratio of the critical currents of the left and right junctions
of the SQUID), which coincides with the minimum of the Peierls-Nabarro
potential.
\end{abstract}
\maketitle

\section{Introduction} \label{intro}

Arrays of Josephson devices have been studied intensively during the 
last several decades \cite{wzso96pd,u98pd}. The recent interest to
these objects has been stimulated by the applications in quantum 
computing \cite{ars06prb,fssk-s07prb} or design of metamaterials,
based on the arrays of rf-biased SQUIDs (superconducting quantum 
interference devices) \cite{lt13sst}. The effect of relativistic time 
dilation has been suggested in the array of asymmetric 
SQUIDs \cite{nknfh10pc}. This system is described by the double 
discrete sine-Gordon (DDbSG) equation \cite{nknfh10pc} and it will be 
the subject of the current article.

On the other hand, the problem of topological soliton (kink or
antikink) mobility in discrete media has attracted much attention 
within recent years. It has been demonstrated that solitary waves 
in nonlinear lattices can propagate with constant shape and velocity 
without any radiation \cite{s79prb,zes97pd,fzk99pre,sze00pd,kzcz02pre,acr03pd,bop05pre,opb06n,dkksh08pre,amp14prl}
despite the presence of the non-zero Peierls-Nabarro potential. 
These solutions are spatially localized (have finite energy) and exist 
for a discrete set of velocities, in contrast to the
continuous models where the allowed velocities usually occupy a continuous
interval, bound from the above by some critical velocity value. Since 
their existence is associated with avoiding the
 resonances with the linear spectrum of the underlying system, they
are called {\it embedded} solitons  \cite{cmyk01pd}. These solitons 
 exist in 
different systems, including the continuum versions of the
sine-Gordon \cite{ck00pre} and double sine-Gordon \cite{bkm01wm} equations
with the high-order dispersion.
In this article it will be shown that the DDbSG equation, that models the 
asymmetric SQUID array, also possesses discrete embedded solitons. 
Taking into account that the model of the SQUID array is 
based on the RCJM (resistively and capacitively shunted
junction model), the consideration of dissipation and external dc bias
is essential, and this also will be done in the current article.

Thus, the aim of this work is twofold: (i) to study the dynamical 
fluxon properties in the SQUID array, and, in particular, 
to obtain the current-voltage characteristics (CVCs) of the array;  (ii)
to demonstrate the signatures of the fluxon as an embedded soliton
that moves along the array without significant radiation.

The paper is organized as follows. The model of the asymmetric SQUID array 
is described in the next section. In Sec. \ref{ham} we discuss the 
properties of the
DDbSG lattice in the Hamiltonian limit. Next section is devoted to the
current-voltage characteristics. Discussion and conclusions are
given in the last section.

\section{The model}\label{model}

\begin{figure}[htb]
\includegraphics[height=0.49\columnwidth,angle=0]{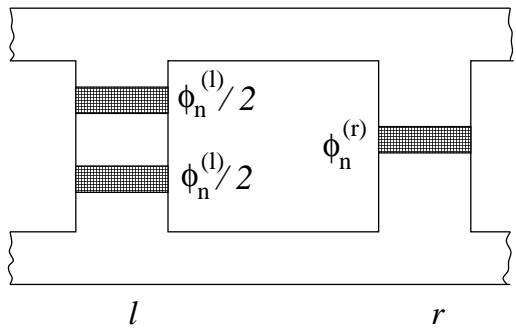}
\caption{Schematic view of the elementary cell of the SQUID
array. The detailed equivalent scheme is given by Fig. 1 
of Ref. \cite{nknfh10pc}. Also see text for details.}
\label{fig0}
\end{figure}

In this paper, we discuss the dynamics of the dc-biased array of 
asymmetric three-junction SQUIDs. Each elementary cell of the array 
is the SQUID that consists of three Josephson junctions, two 
identical junctions are placed in the left arm and one of them
is placed in the right arm, as shown schematically in Fig. \ref{fig0}
(a more detailed equivalent scheme is given in Fig. 1 of 
Ref. \cite{nknfh10pc}).
In the $n$th SQUID $\phi_n^{(l)}/2$ is the Josephson phase
of one of the left junctions and $\phi_n^{(r)}$ is the Josephson
phase of the right junction. 
%
The whole system is described by the dc-driven and damped DDbSG 
equation, which can be written in the dimensionless form as follows
\begin{eqnarray} \nonumber
&&\ddot{\phi}_n - \kappa\, \hat{\Delta} \phi_n +\frac{2}{1+2\eta}\left (
\eta\sin \phi_n + \sin {\phi_n \over 2} \right )+ \alpha  
\dot{\phi}_n=\gamma,\\
\label{1}
&& ~ ~n = 1,2, \ldots,N~.
\end{eqnarray}
The derivation of this equation from the Kirchhoff laws and the Josephson
equations has been performed previously \cite{nknfh10pc}. Under the
assumption of the small loop size the single phase difference
 $\phi_n=\phi_n^{(l)}=\phi_n^{(r)}$ has been introduced
and $\hat{\Delta}\phi_n\doteq \phi_{n+1}-2 \phi_n +\phi_{n-1}$
is the discrete Laplacian. In this model only the self-inductance is
taken into account, while the mutual inductances of the SQUIDs are
 neglected, in accordance with the previous work \cite{bzpo94prb}. 
Other dimensionless parameters are defined as follows
\begin{eqnarray}\label{2}
&&\alpha=\frac{1}{RC\omega_J},~~\kappa=\frac{\Phi_0}{2\pi LI_c},~~ 
\eta=\frac{I_c^{(r)}}{I_c^{(l)}}~,\\
&&C=C_l+{C_r \over 2},~~
\nonumber
{1 \over R}={1 \over R_r}+{1 \over 2R_l},~~ I_c=I_c^{(r)}+
{I_c^{(l)}\over 2}~.
\end{eqnarray}
Here $\omega_J=\sqrt{2eI_c/(C\hbar)}$ is the Josephson plasma frequency 
and the dimensionless time in Eq. (\ref{1}) is normalized in the units 
of $\omega_J^{-1}$, $\alpha$ is 
the dissipation parameter, $\Phi_0$ is the magnetic flux quantum, $L$ 
is the elementary cell inductance and $\gamma$ is the dimensionless external 
bias current, normalized to $I_c$. 
Next, $R_{r,l}$, $C_{r,l}$ 
and $I_c^{(r,l)}$ are, respectively, the resistance, capacitance and  critical 
current of the right or the left junction (marked by the 
sub(super)script ``$r$" or ``$l$"). The parameter $\eta$ measures the asymmetry 
of the SQUID and is the ratio of the critical currents of the right 
and left junctions of the SQUID.

Two limits of Eq. (\ref{1}) are important. If $\eta=0$, one obtains
the discrete sine-Gordon (DSG) equation with the term $\sin (\phi_n/2)$, 
while if $\eta\to \infty$ the DSG equation is restored, but with 
the $\sin \phi_n$ term. The former case physically means that 
the $I_c^{(r)}\to \infty$, thus the left arm of the SQUID effectively disappears.
The latter case means that $I_c^{(l)}\to \infty$ and the same happens
to the right arm. In both the cases the elementary cell of the 
array becomes symmetric.

The circular array is to be considered, thus, the boundary conditions
read $\phi_n=\phi_{n+N}+4\pi Q$, where $Q$ is the total topological charge,
i.e., the total number of fluxons and anti-fluxons trapped in the ring.
In this article only the case of one fluxon in the array will be
considered, hence $Q=1$.

\section{The Hamiltonian limit}\label{ham}

\begin{figure}[htb]
\includegraphics[height=0.8\columnwidth,angle=0]{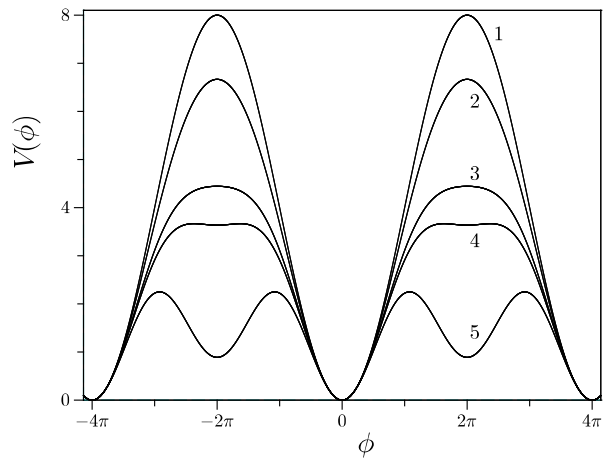}
\caption{Potential Eq. (\ref{4}) for the different values of the
anisotropy parameter: $\eta=0$ (curve 1), $\eta=0.1$ (curve 2),
$\eta=0.4$ (curve 3), $\eta=0.6$ (curve 4) and
$\eta=4$ (curve 5).
 }
\label{fig1}
\end{figure}

The fluxon dynamics in the real SQUID array can be understood better if
 the Hamiltonian limit $\alpha=\gamma=0$ is considered first.
As a result, Eq. (\ref{1}) can be considered as the equation of motion
of the lattice that is governed by the Hamiltonian function
\begin{equation}\label{3}
H=\sum_{n=1}^N \left [ \frac{\dot \phi_n^2}{2}+\frac{\kappa}{2}
(\phi_{n+1}-\phi_n)^2+V(\phi_n)\right ],
\end{equation}
where the on-site potential $V(\phi)$ is expressed as
\begin{eqnarray}\nonumber
V(\phi)&=&V_0\left [\eta (1-\cos \phi)+2 \left (1-\cos \frac{\phi}{2}
\right ) \right ], \\
V_0&=&\frac{2}{1+2\eta}~. \label{4}
\end{eqnarray}
The variable $\phi_n$ can be treated as the coordinate of the respective
particle of the lattice. The shape of the potential (\ref{4}) is
depicted in Fig. \ref{fig1}. It can be clearly seen that the parameter
$\eta$ modifies the shape of the potential significantly. If $\eta=0$
we obtain the sine-Gordon potential with the spatial period $4\pi$.
%
%
For the small values of $\eta$ the potential barrier lowers, and
if $\eta>1/2$ a new local minimum appears at $\phi=2(2n+1)\pi$, $n\in\Z$.
In the limit $\eta\to \infty$ again the sine-Gordon potential is
obtained, however, now its period is $2\pi$. 

The plane waves (Josephson plasmons) can be obtained via linearization of
the equation of motion around the minima of the potential (\ref{4}).
The obtained dispersion laws read
\begin{eqnarray}\label{wl}
\omega_0(q)&=&\sqrt{1+4\kappa~ \sin^2\frac{q}{2} }~,\\
\omega_\pi(q)&=&\sqrt{\frac{2\eta-1}{2\eta+1}+4\kappa~ \sin^2\frac{q}{2} }~.
\end{eqnarray}
The first dispersion law corresponds to the small oscillations around
the global minimum, while the second law makes sense only if $\eta>1/2$
and corresponds to the small oscillations around the metastable state.
Due to finiteness of the array, the wavenumber $q \in [0,2\pi)$ attains
only discrete set of values $q_m=2\pi m/N$, $m=\pm 1,\ldots  , \pm N$.
Note that the dispersion law (\ref{wl}) does not depend on the
asymmetry parameter $\eta$.

\subsection{The continuum limit}\label{continuum}

If $\kappa\gg 1$ the continuum limit can be applied and 
Eq. (\ref{1}) reduces to the double sine-Gordon equation
\begin{equation}\label{6}
\phi_{tt}-\phi_{xx}+\frac{2}{1+2\eta}\left (
\eta\sin \phi + \sin {\phi \over 2} \right )+ \alpha {\phi}_t=\gamma~,
\end{equation}
where the subscripts $_t$ and $_x$ correspond to the time and space 
derivatives, respectively. 
This equation has a large number of applications \cite{cgm83prb,cps86pd},
including the long Josephson junctions with the second harmonic 
in the current-phase relation \cite{gkkb07prb}.

The double sine-Gordon equation has topological soliton solutions 
that connect
two adjacent global minima ($\phi=0$ and $\phi=4\pi$) and in the 
Hamiltonian
limit $\alpha=\gamma=0$ this solution reads 
\cite{cgm83prb,cps86pd,bkm01wm}
\begin{equation}\label{7}
\phi(x,t)=2\pi+4 \arctan \left [ \frac{1}{\sqrt{1+2\eta}} \sinh
\left (\frac{x-vt}{\sqrt{1-v^2}}\right ) \right ]\;.
\end{equation}
If $\eta=1/2$ the soliton solution experiences an inflexion point 
in its center and for the large values of $\eta$ the two $2\pi$
kinks that constitute the solution (\ref{7}) become well separated.

\subsection{Radiationless motion of discrete solitons. 
Sliding velocities}
\label{radiation}

The dynamics of topological solitary waves in the lattices of the 
class, described by the Hamiltonian (\ref{3}) (often referred to as the 
nonlinear Klein-Gordon lattices)  has been well studied (see the 
reviews \cite{fm96ap,bk98pr} and the references therein). The main 
difference in the kink dynamics between the continuous Klein-Gordon 
model and its discrete counterparts is the following fact: the
discreteness significantly obstructs the free soliton propagation,
which is generic for the continuous models.
If one looks for the {\it travelling-wave} solution of the
form $\phi_n(t)=\phi(n-vt)\equiv \phi(z)$ that satisfies the
differential-difference equation with the delay and advance terms
\begin{equation}\label{8}
v^2 \phi''(z)-\kappa \left [\phi(z+1)+\phi(z-1)-2\phi(z) \right ]-
V'[\phi(z)]=0,
\end{equation}
he finds normally kinks that form a coupled state with the 
small-amplitude wave and that state travels with the same velocity
$v$. In the continuum Klein-Gordon models the domain of admissible 
kink velocities is the interval $|v|<1$. Thus, kinks in these continuum 
models form an one-parametric family of solutions with the kink velocity
$v$ being this parameter.

A detailed analytical \cite{s79prb,fzk99pre,opb06n,dkksh08pre,amp14prl} 
and numerical \cite{fzk99pre,sze00pd,kzcz02pre,acr03pd} analysis 
shows that  in a general case of the discrete Klein-Gordon model, the 
continuous family of moving kinks 
is reduced to the discrete finite set of
{\it monotonic} ($\lim_{|n|\to \infty}\phi_n\to const$) 
travelling kink solutions with the velocities
$v=\{v_0\equiv 0, v_1,v_2 \ldots ,v_k\}$. Further on, all velocities that
satisfy $v_ n\neq 0$ will be called {\it sliding} velocities since the kink
slides along the lattice with these velocities without any radiation.
These solutions can be called {\it discrete embedded solitons}.
In the DSG equation there is only non-mobile ($v_0=0$) monotonic kink and
 there is no sliding velocities. 
In general, everywhere away from the sliding velocities, i.e., if $v\neq v_n$,
the moving kinks are non-monotonic, have oscillating asymptotic tails 
and are often referred to as {\it nanopterons}. The monotonic solutions 
 are of big importance since their energy is finite.

\begin{figure}[htb]
\includegraphics[height=0.7\columnwidth,angle=0]{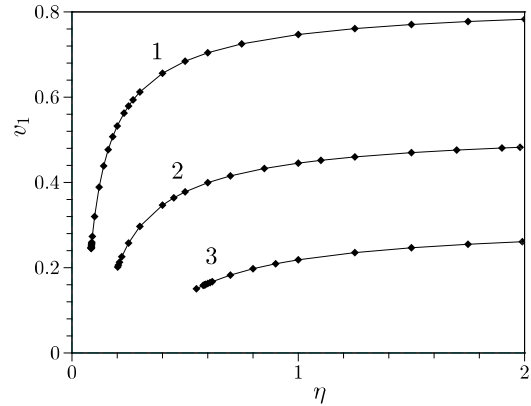}
\caption{Sliding velocities as a function of $\eta$ for $\kappa=1$ 
(curve $1$), $\kappa=0.5$ (curve $2$) and $\kappa=0.25$ (curve $3$).
The solid lines are used as a guide for an eye.
 }
\label{fig2}
\end{figure}

Using the so-called {\it pseudospectral} method, developed 
in \cite{hmb88pd,ef90pla,defw93pd}, it is possible to compute the solution 
 of Eq. (\ref{8}) with the arbitrary desired numerical precision. In 
order to compute the {\it monotonic} solitary wave one has to monitor 
the tail of the solution $\phi(z)$ and change the velocity $v$ until the
amplitude of the oscillating tail becomes smaller than the defined 
tolerance value.  
We have done that for the DDbSG equation and in Fig. \ref{fig2} we 
%
%
show the dependence of the first sliding velocity, $v_1$, on the asymmetry
parameter $\eta$ for the different values of the discreteness 
parameter $\kappa$. For these sets of parameters the spectrum of sliding velocities consisted of only one velocity, $v_1$. It appears that even 
for the rather small values of $\eta$ (even $\eta\ll 1$, provided
$\kappa \gtrsim 1$) there exists at least one sliding velocity.
The value of the sliding velocity is smaller for the smaller values
of the coupling parameter, and, 
it is interesting that the monotonic moving kinks can exist even in the
strongly discrete lattice with $\kappa=0.25$. The dependence
$v_1(\eta)$ starts from some critical value of $\eta$, below this 
value the system does not allow for the sliding velocities, this is in
agreement with the general theory of \cite{acr03pd}. In the
quasicontinuum approximation of DSG \cite{ck00pre} or DDbSG \cite{bkm01wm}, 
when the discrete
Laplacian is approximated up to $\phi_{xxxx}$, the moving kink exists
again only for the selected set of sliding velocities, but these
velocities (depending on the model parameters) can be arbitrarily small.

\begin{figure}[htb]
\includegraphics[width=1.05\columnwidth,angle=0]{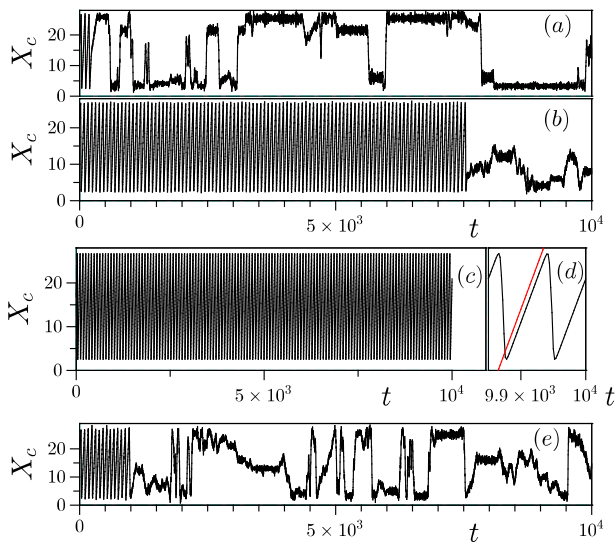}
\caption{(Color online) Kink center of mass evolution in the
Hamiltonian limit 
for $\kappa=0.5$, $\eta=0.6$, $v_1=0.399493$ 
with the initial boost velocity
$v_*=0.97v_1$ (a), $v_*=0.99v_1$ (b), $v_*=v_1$ (c,d) and $v_*=1.25v_1$ (e).
The red line in the panel (d) corresponds to the center of mass
moving with the sliding velocity $X(t) \sim v_1 t$.
 }
\label{fig3}
\end{figure}

The substitution of the solution of Eq. (\ref{8}) with the sliding velocity 
$v_1$ into the actual DDbSG equation (\ref{1}) results in the kink 
propagation 
continuously with this velocity $v_1$ without any noticeable radiation.
The following numerical experiment demonstrated this. The moving
kink obtained as a solution of Eq. (\ref{8}) with the sliding
velocity $v_1$ has been launched with the
velocity $v_*$, that may differ from $v_1$. The evolution of the 
soliton center of mass is shown in Fig. \ref{fig3}. 
%
%
It appears that if the initial soliton velocity differs from $v_1$ 
insignificantly
the soliton can travel for a long time around the 
lattice [see Fig. \ref{fig3}(b)]. Otherwise
it gets pinned rather fast as shown in Figs. \ref{fig3}(a,e).
If $v_*=v_1$ the soliton travels without any significant slowing down
or radiation [see the panels (c) and (d)].

\section{The current-voltage characteristics}\label{cvc}

Now we can start constructing the current-voltage characteristics
(CVCs) for the non-zero values of bias and dissipation.

\subsection{The continuum limit}\label{cvc1}

In the continuum 
limit one can use the power balance method from Ref. \cite{ms78pra}
and compute the equilibrium fluxon velocity. The power balance equation 
reads $\bar{V}_c \gamma = 4\pi\sqrt{\kappa}~ v \gamma/N=-P_{diss}$. 
Here $\bar{V}_c$ is the average voltage drop, produced by the fluxon
moving with the velocity $v$. If we assume that the perturbation,
caused by the bias and dissipation, is small and the fluxon shape is
given by the exact solution  of the unperturbed continuous
equation (\ref{6}), we can compute the power of the dissipative losses
\begin{eqnarray}
&&P_{diss}=-\alpha \int_{-\infty}^{+\infty}\phi_t^2 dx=
-\frac{16 \alpha v^2}{\sqrt{1-v^2}}\Phi(\eta)~, \\
&&\Phi(\eta)= 1+\frac{1}{\sqrt{2\eta(2\eta+1)}}~
\mbox{arctanh} \sqrt{\frac{2\eta}{1+2\eta}},\nonumber
\end{eqnarray}
where the solution (\ref{7}) has been substituted in the above integral. 
From this equation one can find the equilibrium fluxon velocity $v$ 
and the average voltage drop
\begin{equation}\label{11}
\bar{V}_c=\sqrt{\kappa}\frac{4 \pi v_\infty}{N}
=\sqrt{\kappa}\frac{4 \pi}{N}\left [1+ \Phi^2(\eta)
\left ({4\alpha} \over {\pi\gamma}\right)^2 \right ]^{-1/2}~.
\end{equation}
The auxiliary function $\Phi(\eta)$ attains two important limits:  
$\lim_{\eta\to 0}\Phi(\eta)=2$ and $\lim_{\eta\to \infty}\Phi(\eta)=1$. 
In both these limits the well-known formula for the SG 
equation \cite{ms78pra} is
restored. Since $\Phi(\eta)>1$ for any finite positive $\eta$, the
slope of the CVC near the origin ($\gamma \ll 1$) will be more and
more flat as $\eta$ increases: 
${\bar V}_c \simeq \sqrt{\kappa}\pi^2 \gamma /[N\alpha {\Phi(\eta)}]$.

\subsection{The numerical results}\label{numres}

The numerically-computed CVCs are shown in Figs. \ref{fig4}-\ref{fig6}
by the markers while the blue solid lines correspond to the analytical
formula (\ref{11}) that stems from the continuum approximation. 
This approximation predicts the continuous curve $\gamma=\gamma(\bar V)$ 
and appears to work well only for the small values of $\gamma$. 
Surprisingly, it gives the correct slope of the CVC near
the origin even for the strongly discrete array ($\kappa=0.5$), however 
it  works rather poorly for the larger values of the external bias.

The numerically computed average voltage drop is defined as
\begin{equation}\label{12}
 {\bar V}= \frac{1}{N}\sum_{n=1}^N \lim_{t
\rightarrow \infty} \frac{1}{t}\int_0^t {\dot \phi}_n(t') dt'~.
\end{equation}
If the fluxon propagates with the constant velocity $v$ it produces 
the average voltage drop ${\bar V}=4\pi v/N$.
The CVC calculation procedure can be described as follows.
We start at the zero bias ($\gamma=0$) and integrate 
the equations of motion (\ref{1}) with the 4th
order Runge-Kutta method for each value of $\gamma$ during the time
$t > 10 \alpha^{-1}$. When the average voltage (\ref{12}) reaches the 
desirable
tolerance, $\gamma$ is increased by some small amount and the procedure
is repeated again. The same calculation has been performed when $\gamma$
is decreased till $\gamma=0$. Since the CVCs demonstrate complex 
hysteretic structure (sometimes multiple), each branch has been 
path-followed back and forth between its respective ends.

The typical CVCs consists of the cascades
of separate branches that appear due to the fluxon
coupling with the plasmon modes, see theoretical 
\cite{zows95prl,wzso96pd,ucm93prb,bhz00pre} and experimental
papers \cite{zows95prl,wzso96pd,clumol93pla,uclocr95prb}.
While moving along the array, the fluxon excites the plasmon modes
and forms a bound state that propagates with the same velocity. If 
$v$ is the fluxon velocity, the plasmon phase velocity should equal $v$
as well. Thus, the plasmon wavenumber is given by the root of the
following equation:
\begin{equation}\label{11a}
\omega_0(q)-vq=0~,
\end{equation}
where $\omega_0(q)$ is the plasmon dispersion law (\ref{wl}).
Due to periodicity of the boundary conditions the phase locking
in the array would occur if the finite number of the Josephson phase 
oscillations
will fit into one cycle of the fluxon journey along the array.
In other terms
this means that a certain number of the plasmon wavelength should be fitted
in the array, as one can easily see from the inset in Fig.\ref{fig4}(a),
where the $\dot \phi_n$ distribution is given from the two neighbouring
voltage steps. Thus, the different number of oscillations fitted in the
array corresponds to the different branch of the CVC.
This situation has been reported in the 
literature \cite{ucm93prb,wzso96pd} together with the approximate 
values of the voltage steps, so we will not dwell on it any longer.

\subsubsection{Signatures of the sliding velocities}

Analysis of these CVCs shows that there exist both qualitative and
quantitative differences between the cases when the DDbSG equation 
has at least one sliding velocity in the Hamiltonian limit and when 
there is no such a velocity. These differences can be summarized in the  
following two paragraphs.

\paragraph[(i)]{\bf The inaccessible interval (gap) within the 
range of admissible voltages.}

\begin{figure}[htb]
\includegraphics[width=1.07\columnwidth,angle=0]{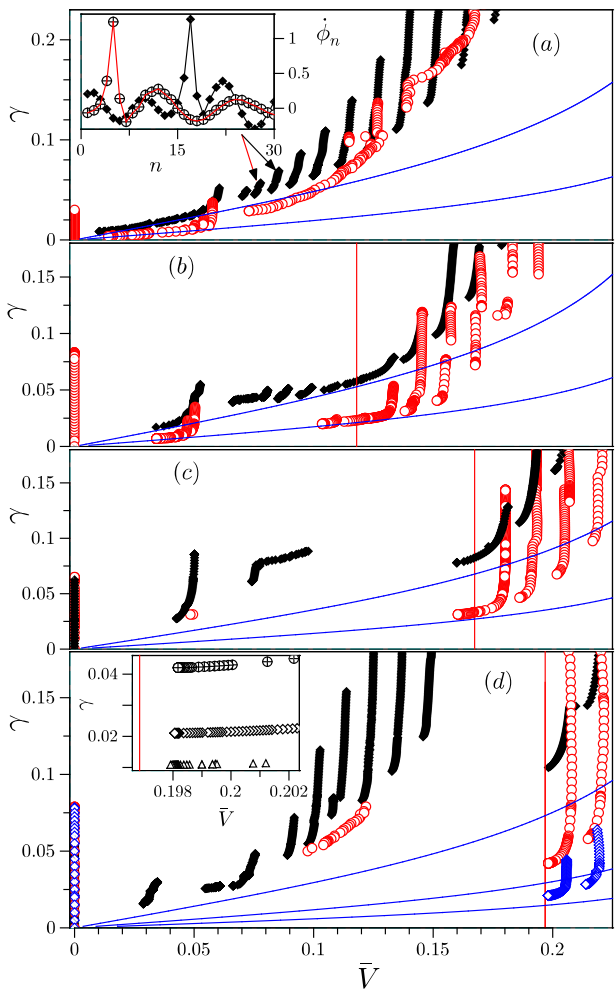}
\caption{(Color online) 
Current-voltage curves for $\kappa=0.5$, $N=30$, $\alpha=0.05$
(black $\blacklozenge$), $\alpha=0.02$ (red $\circ$), $\alpha=0.01$ (blue 
$\diamond$) and $\eta=0.1$ (a), 
$\eta=0.3$ (b), $\eta=0.6$ (c) and $\eta=1.5$ (d). The blue solid lines 
correspond to the respective CVC
in the continuum limit, Eq. (\ref{11a}). The red vertical lines in (b)-(d) 
 are given by $4\pi v_1/N$, where $v_1$ is the sliding 
velocity for the respective value
of $\eta$ (see Fig. \ref{fig2}). The inset in the panel (a) shows the 
distribution 
of ${\dot \phi}_n$ that correspond to the branches of the CVC 
pointed by the arrows at $\alpha=0.05$. The distribution corresponding
to the branch on the right is given by $\blacklozenge$ while 
$\oplus$ corresponds to the branch on the left.
The inset in the panel (d) shows the details of CVCs in the
neighbourhood of the sliding velocity for $\alpha=0.02$ ($\oplus$), 
$\alpha=0.01$ ($\diamond$) and $\alpha=0.005$ ($\Delta$). 
 }
\label{fig4}
\end{figure}

It appears that if $\eta$ is too small, the underlying Hamiltonian 
problem does not possess a sliding velocity, we face the situation 
when the CVC consists of the branches shown in of Fig. \ref{fig4}(a), 
and the interval between these branches along the ${\bar V}$ axis 
decreases as ${\bar V}$ decreases. The same is true about the length 
of these branches in the $\gamma$ direction. 
A small gap between the voltage steps can be noticed in 
 the interval 
$0.06\lesssim {\bar V} \lesssim 0.07$. This interval corresponds to the
situation when the number of roots of Eq. (\ref{11a}) has changed from
one to three. The kind of picture described in Fig. \ref{fig4}(a) is 
the same as observed in the simple driven and damped DSG equation
\cite{ucm93prb}.

Yet a different situation occurs if in the Hamiltonian limit
there is at least one sliding velocity. According to the results
of Subsec. \ref{radiation} for $\kappa=0.5$ there should be a 
sliding velocity if $\eta>0.202$, thus, this is the 
case for the CVCs in the panels (b)-(d). In these figures one can easily
spot a significant {\it inaccessible voltage interval} (IVI) or a gap, 
i.e., there is an forbidden 
interval in ${\bar V} \in [V_{IVI}^-,V_{IVI}^+]$ where no voltage can be 
produced by the moving fluxon.
This interval is not noticeable for $\eta=0.3$, $\alpha=0.05$ 
[see Fig. \ref{fig4}(b)], but is clearly seen for $\alpha=0.02$. 
However, the IVI 
increases strongly as  $\eta$ increases [see Figs. \ref{fig4}(c,d)].
It is important to remark that the upper edge of this inaccessible
interval, $V_{IVI}^+$, moves closer
and closer towards the voltage $4\pi v_1/N$, produced the fluxon 
moving with the sliding velocity $v_1$. 
%
%
\paragraph[(ii)]{\bf The IVI 
increases as the damping parameter decreases.}

\begin{figure}[htb]
\includegraphics[width=1.09\columnwidth,angle=0]{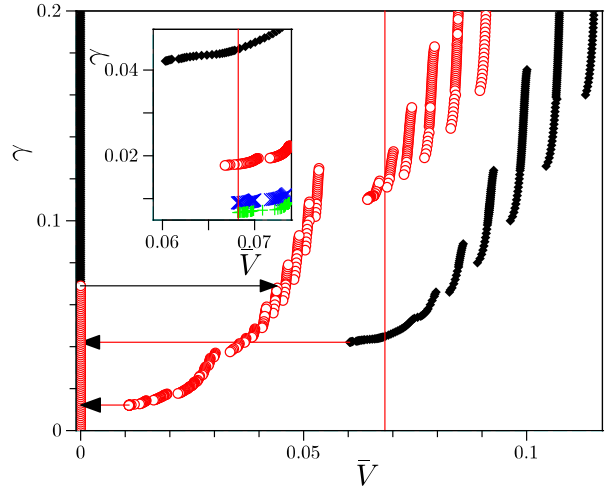}
\caption{(Color online) Current-voltage curves for $\kappa=0.25$, $\alpha=0.05$, 
$\eta=0.6$ (black $\blacklozenge$) and $\eta=0.3$ (red $\circ$). 
The inset  shows the details of CVCs in the
neighbourhood of the sliding velocity for $\eta=0.6$,
$\alpha=0.05$ (black $\blacklozenge$), 
$\alpha=0.02$  (red $\circ$) 
$\alpha=0.01$ (blue $\times$) and 
$\alpha=0.0075$ (green $+$). The red vertical line marks the voltage drop 
that corresponds to the respective 
sliding velocity in the Hamiltonian limit.
 }
\label{fig6}
\end{figure}

Another important observation is that as the dissipation decreases, 
the width of the IVI increases. The two sets of data are 
plotted in Figs. \ref{fig4}(b-c),
for $\alpha=0.05$ and $\alpha=0.02$. In addition, in Fig. \ref{fig4}(d) 
the data for $\alpha=0.01$ are given. 
One can notice that the IVI becomes quite pronounced for $\eta=0.3$ if
the damping coefficient is reduced from $\alpha=0.05$ to $\alpha=0.02$.
This is seen even better in Figs. \ref{fig4}(c,d) where the IVI is
well defined for $\alpha=0.05$ and its width increases with the
growth of $\eta$ and with the decreasing of $\alpha$.
In particular, we note that there are fewer branches for ${\bar V}$ below
the IVI (${\bar V} < V_{IVI}^-$) if $\alpha$ is decreased. 
The length of these branches along the $\gamma$ axis decreases as well, 
compare, for example, the data in Fig. \ref{fig4}(d), where only one 
branch below the IVI survives 
if the dissipation parameter is reduced from $\alpha=0.05$ to 
$\alpha=0.02$. If $\alpha$
is reduced further till $\alpha=0.01$ there is no other CVC branches below
the IVI, i.e., $V_{IVI}^-=0$. The two CVCs for the smaller value of 
the discreteness constant 
($\kappa=0.25$) for $\eta=0.3$ (no sliding velocity) and
 $\eta=0.6$ (one sliding velocity $v_1=0.162871$) are
given in Fig. \ref{fig6}.  
%
%
Here the IVI is clearly seen for $\eta=0.6$ and the behaviour of
its edges is the same as in Fig. \ref{fig4}: $V_{IVI}^+ \to 4\pi v_1/N$
as $\alpha \to 0$ (see the inset) while $V_{IVI}^-=0$.

The upper edge of the IVI, $V_{IVI}^+$ is positioned close to the
value ${4\pi}v_1/N$, where $v_1$ is the respective sliding
velocity. By defining the detuning parameter
\begin{equation}\label{14}
\nu=\left |V_{IVI}^+-\frac{4\pi}{N}v_1\right |~,
\end{equation}
and presenting it in Tab. \ref{tb1}, we demonstrate 
that the IVI is directly associated with the sliding velocity of
the DDbSG equation in the Hamiltonian limit.
\begin{table}[htb]
\begin{tabular}{cccc}
\hline
\hline
$\alpha \backslash \eta$ & 0.6      &  1  &  1.5 \\
\hline      
  0.05               & 0.0075 & 0.0017 &  0.0013     \\
  0.02               & 0.0070 & 0.0014 &  0.0013      \\
  0.01               & 0.0050 & 0.0014 &  0.0012     \\
  0.005              & 0.0041 & 0.0010 &  0.0011      \\
\hline  
\hline
\end{tabular}
\caption{The detuning parameter $\nu$ [see Eq. (\ref{14})] for 
$\kappa=0.5$ as a function of $\alpha$ and $\eta$.}
\label{tb1}
\end{table}
Indeed, the detuning parameter decreases as $\eta$ increases as well
as $\alpha\to 0$.

These results can have the following mathematical interpretation. If the 
Hamiltonian system possesses a sliding velocity $v$, even the small
perturbation by adding non-zero $\alpha$ and $\gamma$ creates an
attractor that corresponds to the fluxon motion with the velocity close
to $v$ (see Ref. \cite{kz04nova}). Thus, for small $\alpha$ we observe 
that as
 $\gamma \to 0$, the fluxon moves with the velocity close to $v$.  
If there is no sliding velocity, for the same values of $\alpha$ 
and $\gamma$ the fluxon either moves with very small velocity close to $0$,
or is simply pinned due to discreteness. This situation is well seen
in Figs. \ref{fig4}(a) and \ref{fig6}, where the CVC approaches close
to the origin but never attains it.

\subsubsection{Chaotic vs. regular regimes of motion}

\begin{figure}[htb]
\includegraphics[width=1.0\columnwidth,angle=0]{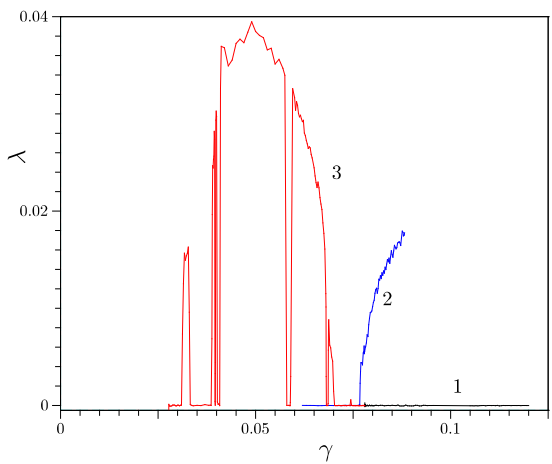}
\caption{(Color online)
Largest Lyapunov exponent as a function of bias for three
branches of the CVC, given in Fig. \ref{fig4}(c). The parameters are:  
 $\kappa=0.5$, $\alpha=0.05$ and $\eta=0.6$. Different lines 
 correspond to the different branches of the CVC. See text for details. }
\label{fig7a}
\end{figure}

Observation of the branches of on the CVCs in Figs. \ref{fig4}-\ref{fig6}
reveals that the most of these branches are almost vertical lines.
Only at their bottom ends 
these branches become bent towards the lower voltages. The branches
near the IVI are almost horizontal with small vertical parts. This
transition goes on smoothly as $\gamma$ decreases. However,
there are some isolated branches that fall out from the usual picture. 
For example, in Fig. \ref{fig4}(c) there is a branch just below the IVI, 
with the weakly tilted top part and almost vertical bottom part.
Also, in Fig. \ref{fig4}(d) at $\alpha=0.02$ there is an isolated 
and significantly tilted branch just below the IVI. 
Hence, one should focus on the nature of the dynamics that fluxon 
undergoes when traversing the array. 
In order to solve this question, the largest Lyapunov exponent 
(LLE) $\lambda$ has been computed (with the help of the 
Benettin algorithm \cite{ben}) for the three branches of 
the CVC from Fig. \ref{fig4}(c) at $\alpha=0.05$. We have taken 
the closest branches to the IVI, one above it, with voltages changing in 
the range $0.16 \lesssim {\bar V}\lesssim 0.18$, and two below it, with the
voltages in the range $0.074\lesssim {\bar V}\lesssim 0.1$ and 
$0.043 \lesssim {\bar V}\lesssim 0.05$. The respective dependencies $\lambda(\gamma)$ are given by the lines
1 (black), 2 (blue) and 3 (red) in Fig. \ref{fig7a}(a). We observe that 
$\lambda=0$ for the line 1 (black), and this line corresponds to 
the branch of the CVC that is just above the IVI. For the  
non-bounded trajectories in the autonomous system there is always
a zero Lyapunov exponent, thus the dynamics on this branch is regular.
In the curve 2 (blue) LLE is positive in the interval 
$0.077 \lesssim\gamma \lesssim 0.088$ that corresponds approximately 
to the top part of the branch and becomes zero in the interval 
$0.062\lesssim \gamma\lesssim 0.077$ that corresponds approximately to 
the bottom part of the branch. From the respective CVC [Fig. \ref{fig4}(c)] 
one can determine that the dynamics is chaotic when the branch is strongly 
tilted and is regular when the CVC is almost vertical.
%
Finally, the dependence $3$ for the lowest (closest to the origin) 
branch shows that dynamics is regular at the upper part of the branch and
becomes chaotic as $\gamma$ decreases. Several switchings from chaotic to
regular motion and back can be observed as well.

\begin{figure}[htb]
\includegraphics[width=1.08\columnwidth,angle=0]{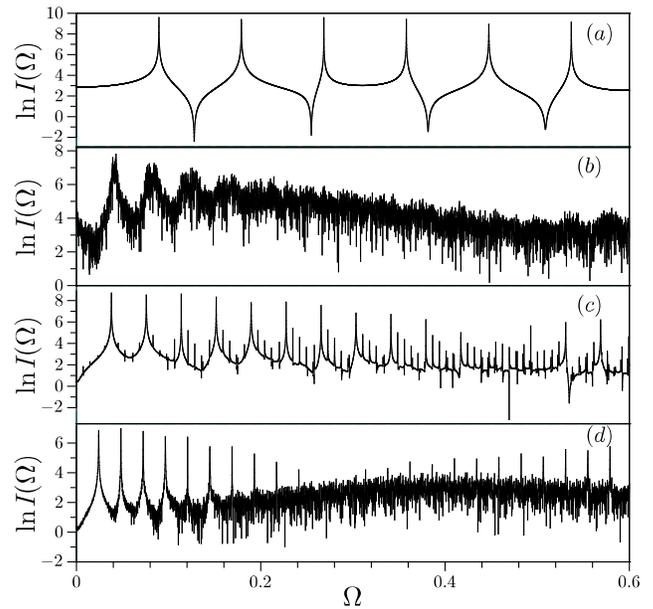}
\caption{
Power spectra of the ${\dot \phi}_{N/2}$ variable 
for the different values of bias at 
(a) $\gamma=0.11$, ${\bar V}=0.1789$; 
(b) $\gamma=0.08$, ${\bar V}=0.08241$;
(c) $\gamma=0.075$, ${\bar V}=0.07585$;
(d) $\gamma=0.045$, ${\bar V}=0.04825$.
 }
\label{fig7b}
\end{figure}

The power spectrum of ${\dot \phi}_{N/2}$, defined as
\begin{equation}
I(\Omega)=\left |\int_{-\infty}^{+\infty} 
{\dot \phi}_{N/2}(t) e^{-i\Omega t} dt\right |^2~,
\end{equation}
is plotted in Fig. \ref{fig7b}. It has been computed for the several 
values of $\gamma$ at the CVC branches, discussed in the previous paragraph. 
First we consider 
the branch above the IVI ($\bar V > V_{IVI}^+$), for which the LLE
is always  zero (line 1 in Fig. \ref{fig7a}). As one can see in 
Fig. \ref{fig7b}(a), the spectrum consists of the equidistant peaks, 
positioned at $\Omega=n {\bar V}/2$, $n=1,2,\ldots$. Thus, the trajectory
is the limit cycle with the frequency ${\bar V}/2$. This 
means that the fluxon reconstructs its shape completely 
after travelling around the array twice.
The respective position on the CVC [see Fig. \ref{fig4}(c)]
corresponds to $\gamma=0.11$, ${\bar V}=0.1789$.

Next we turn our attention to the second branch of the CVC, that lies 
just below the IVI. On the weakly tilted part of the branch 
($\gamma=0.08$) the dynamics is chaotic, as has been seen from the
LLE dependence (line 2 in Fig. \ref{fig7a}). The power spectrum consists
of the wideband [see Fig. \ref{fig7a}(b)] and several peaks at 
$\Omega\sim {\bar V}/2, {\bar V}, 3{\bar V}/2$. The average voltage drop
here is $\bar V =0.0825$. Another point on the same branch
corresponds to the vertical part of it ($\gamma=0.075$). Dynamics there is
regular, the peaks [see the panel (c)] 
at $\Omega=n {\bar V}/2$, $n=1,2,\ldots$ can be easily spotted  
 in the low-frequency region. There are other peaks, associated with
 some frequency that is significantly lower than ${\bar V}$. Hence, the
 respective trajectory is quasiperiodic.
%
Finally, on the closest to the origin branch of the CVC at $\gamma=0.045$
chaotic dynamics has been observed (see line 3 of Fig. \ref{fig7a}).
This is confirmed further by the broadband power spectrum, 
shown in Fig. \ref{fig7b}(d). However, the peaks at 
$\Omega=n {\bar V}/2$, $n=1,2,\ldots$ can be spotted much
better comparing to 
another chaotic case, shown in the panel (b).
We have also checked the dynamics of the isolated branch in 
Fig. \ref{fig4}(d) at $\alpha=0.02$. Both the LLE calculations and the 
spectral analysis
show that the dynamics there is also chaotic and the power spectrum
is similar to the spectrum in Fig. \ref{fig7b}(b).

\subsubsection{The critical depinning current}

\begin{figure}[htb]
\includegraphics[width=1.01\columnwidth,angle=0]{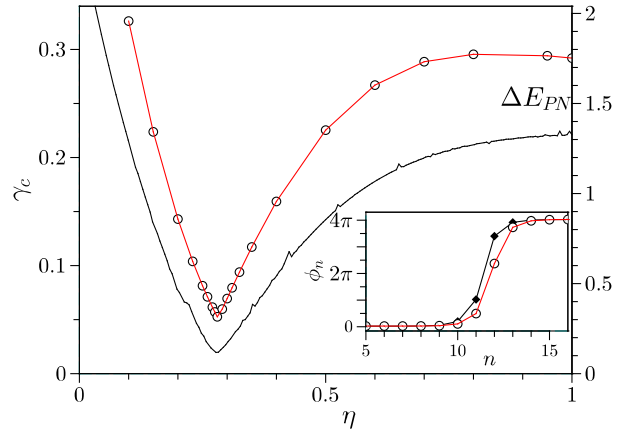}
\caption{(Color online). Critical depinning current ($\circ$, left scale)
and the PN barrier (solid black line, right scale) as a function of 
the asymmetry parameter $\eta$ for 
$\kappa=0.25$. The red line is used as a guide for an eye.
The inset shows the distribution of the $\phi_n$ for the
pinned fluxon at $\eta=0.26$, $\gamma=0.07111$ ($\blacklozenge$) and
$\eta=0.31$, $\gamma=0.079$ ($\circ$). The solid lines are used as
guides for an eye.}
\label{fig7}
\end{figure}

The careful investigation of the CVCs demonstrates the non-monotonic 
dependence of the
critical depinning current $\gamma_c$ on the asymmetry parameter $\eta$.
The critical depinning current is the minimal bias
current which can sustain the pinned fluxon state ($\bar V=0$ on the CVC).
Indeed, as on can see from Fig. \ref{fig6}, the critical depinning current
equals $\gamma_c=0.0695$ for $\eta=0.3$ and $\gamma_c=0.267$ for $\eta=0.6$.
At first glance this seems to be surprising, as we can naively
suppose that the pinning of the fluxon is defined by the barrier height
of the double sine-Gordon potential $V(\phi)$ (\ref{4}). This suggestion 
is obviously 
wrong, since the height of $V(\phi)$ decreases with the growth of $\eta$.
Moreover, the dependence of the critical current $\gamma_c$ on the asymmetry
parameter $\eta$ appears to be non-monotonic, as shown in Fig. \ref{fig7}.
In order explain this behaviour it is useful to compute the PN 
potential and its barrier as a function of $\eta$. The concept
of the PN potential is known for a long time \cite{bk98pr} and is used 
to describe the motion a topological soliton in the discrete 
media as
a motion of an inertial particle in the field produced by the spatially
periodic potential. This is, in fact, the PN potential, $V_{PN}(X)=V_{PN}(X+1)$
and $X$ is the soliton center of mass.
The PN barrier is defined as $\Delta E_{PN}=\max_{X} [V_{PN}(X)]-\min_X [V_{PN}(X)]$.
If $\kappa \gg 1$ the PN potential can be computed analytically\cite{im82jpsj}
and for the DSG equation it satisfies $V_{PN}(X) \propto 1-\cos {X}$. 
In our case the 
perturbational approach will fail, therefore the PN potential has to 
be computed numerically.  

In Fig. \ref{fig7} the PN barrier as a function of the asymmetry 
parameter $\eta$ is demonstrated by the solid black line. The dependence 
is non-monotonic
and has a clear minimum for the same $\eta$ where the $\gamma_c(\eta)$
has a minimum. At this minimum the barrier is decreased by the order
of magnitude.
%
This non-monotonic behaviour of the PN barrier as a function
of the shape parameter (and also of the coupling constant $\kappa$) 
is already known for other discrete Klein-Gordon
systems \cite{pr82prb,sze00pd,kzcz02pre} with the on-site potentials
that depend on the shape parameter. The reason is the following.
Under the certain set of parameters $\Delta E_{PN}$ both the
site-centered and bond-centered static kink states are the 
local minima of the total 
energy, while the local maximum corresponds to the asymmetric 
configuration, which is intermediate between the site-centered
and bond-centered states. 
For these parameters the function $\Delta E_{PN}$ attains its 
minimum. It is important to note that if $\eta=0$, i.e., in the
case of the conventional DSG equation, the bond-centered state is always
a local minimum and the site-centered state is always a saddle 
point. Thus, for the DSG equation $\Delta E_{PN}$ is simply the 
difference of the energies of these kink configurations.
The inset of Fig. \ref{fig7} shows the pinned fluxon profiles for
$\eta=0.26$  and $\eta=0.31$, while the $\gamma_c(\eta)$
dependence attains its minimum at $\eta \approx 0.28$. In the 
former case the bond-centered fluxon is the local minimum of the
energy while in the latter case it is the site-centered fluxon. 
The fluxon profiles look a bit asymmetric because  the non-zero
bias makes the total potential energy asymmetric with respect to the
point $\phi=0$.

%
\section{Discussion and conclusions}

In this article it has been demonstrated how the features of the kink 
mobility in the discrete Klein-Gordon models can be manifested in 
realistic systems. As a particular system the array of the asymmetric 
three-junction SQUIDs has been considered. This object is described by 
the discrete double sine-Gordon equation (DDbSG). 

The main result can be summarized as follows. In the Hamiltonian limit
the DDbSG equation alongside with other similar models, like the
Peyrard-Remoissenet \cite{pr82prb} and the double Morse \cite{kzcz02pre}
chains, allows for a discrete set of kink velocities ({\it sliding} 
velocities) with which monotonic 
($\lim_{|n|\to \infty}\phi_n \to const$) kinks can propagate.
These excitations belong to the family of the so-called {\it embedded}
solitons \cite{ck00pre,cmyk01pd}.
The signature of the sliding velocities can be spotted on the CVCs of the
array. This signature is a significant inaccessible voltage interval (IVI), 
i.e. the voltage that cannot be produced by the moving fluxon.
As the voltage drop is proportional to the fluxon velocity, one
can speak also about the inaccessible velocity interval.
This interval does not appear if the asymmetry 
parameter $\eta={I_c^{(r)}}/{I_c^{(l)}}$ is too small and there is no 
sliding soliton velocities in the Hamiltonian limit. The IVI becomes more 
pronounced if $\eta$ increases
or if $\alpha \to 0$. In particular, the lower edge of the IVI tends to
zero, while the upper edge converges to the value $4\pi v/N$ with
$v$ being the sliding velocity.

Another important result is the significant 
lowering of the critical pinning current due to the change of $\eta$. 
The explanation is based on the non-trivial dependence of the PN barrier 
on the asymmetry parameter. Similar results on
the lowering of the activation barrier for  solitons  
have been reported earlier for other lattice models
 \cite{pr82prb,cgm83prb,sz91pra}.

It is also important to comment on the connection with the problem
of the radiationless motion of the bunched kink (fluxon) states in the
ordinary DSG lattice. 
The symmetric  SQUID array, or, equivalently the array of parallel shunted
small Josephson junctions \cite{ums98prb} is described by the DSG equation. 
It has been shown in several cases both 
theoretically \cite{pk84pd,ums98prb,sze00pd,acr03pd} and 
experimentally \cite{psau06prl} that the radiationless sliding of the coupled pair
of several kinks ($4\pi$, $6\pi$, etc.) is possible for the selected set
of kink velocities. This phenomenon has been treated analytically in the
quasi-continuum approximation in Refs. \cite{ck00pre,bkm01wm}, but it
takes place even in the sufficiently discrete  array ($\kappa<1$) as well.
In the limit $\eta\to \infty$ the double sine-Gordon potential in (\ref{4})
becomes the ordinary sine-Gordon potential with the period $2\pi$, thus, the
above-mentioned result of the bound state of two $2\pi$ kinks is the 
special case of the kink mobility of the 
DDbSG equation in the limit $\eta\to\infty$. 

In the current model the role of the mutual inductances of the array 
cells has been neglected in accordance the previous work \cite{bzpo94prb}. 
If the mutual inductances
 are taken into account, the dynamics of the Josephson
phases should be described not by the DDbSG equation (\ref{1}) but
 by Eq. (1) of Ref. \cite{bzpo94prb}. The main difference between
these two equations lies
in the nature of the coupling term. While in Eq. (\ref{1}) there is 
coupling only between the nearest neighbouring Josephson phases, the case
with mutual inductances accounts for coupling of all Josephson phases
of the array.
The existence of the sliding velocities depends primarily on the properties
of the current-phase relation that contains the $\sin \phi_n$ and the 
$\sin (\phi_n/2)$ terms, and not on the interaction. Therefore, we do not
expect any qualitative differences if the mutual inductances are taken into
account. The quantitative differences may occur because the 
value of the sliding velocity depends of the coupling parameter.

Finally, we remark that the DDbSG equation (\ref{1}) can describe 
another system - the parallel array of Josephson junctions that have 
the biharmonic current-phase relation 
$I_c(\phi)=I_{c,1}\sin \phi +I_{c,2}\sin 2\phi$ \cite{gki04rmp} [here
the substitution $\phi \to \phi/2$ should be performed in order
to get Eq. (\ref{1})]. This, in particular, is true for the  
superconductor-ferromagnet-superconductor (SFS) and
superconductor-ferromagnet-insulator-superconductor (SFIS) junctions.
Naturally, the phenomena, discussed in this paper apply to the 
arrays of the SFS and SFIS junctions as well.

\section*{Acknowledgements}

One of the authors (Y.Z.) acknowledges the financial support from the
Ukrainian State Grant for Fundamental Research No.~ 0112U000056.


\end{document}